\documentclass[twocolumn,preprintnumbers,amsmath,amssymb]{revtex4}
\usepackage{graphicx}
\usepackage{dcolumn}
\usepackage{bm}
\usepackage{mathptmx,amsmath,multirow}
\usepackage[colorlinks, linkcolor=blue, citecolor=blue, urlcolor=blue]{hyperref}
\usepackage{color}
\begin{document}
	\title{Attosecond pulse synthesis from high-order harmonic generation in intense squeezed light\\
	\footnotetext{$^{*}$xylai@wipm.ac.cn\\$^{\dagger}$xjliu@wipm.ac.cn}}
	\author{ShiJun Wang$^{1,2}$, XuanYang Lai$^{1,3*}$, and XiaoJun Liu$^{1\dagger}$}
	\affiliation{$^{1}$State Key Laboratory of Magnetic Resonance and Atomic and Molecular Physics, Wuhan Institute of Physics and Mathematics, Innovation Academy for Precision Measurement Science and Technology, Chinese Academy of Sciences, Wuhan 430071, China\\ $^{2}$School of Physical Sciences, University of Chinese Academy of Sciences, Beijing 100049, China\\ $^{3}$Wuhan Institute of Quantum Technology, Wuhan, 430206, China}
\date{\today}	
	
\begin{abstract}

High-order harmonic generation (HHG) provides a broad spectral bandwidth for synthesizing attosecond pulses. However, in the current HHG schemes, only part of the harmonics can be phase-locked, which limits the ability to achieve shorter attosecond pulses. Here, we study attosecond pulse synthesis from HHG of an atom driven by an intense quantum light, i.e., squeezed light. It is interestingly found that the harmonics in the whole spectrum can be phase-locked and, by using these harmonics, the width of the synthesized attosecond pulse is greatly reduced. By developing strong-field approximation theory in squeezed light, the physics of the phase-locked harmonic generation throughout the HHG spectrum is revealed and is found to be independent of the target system. Furthermore, we uncover the dependence of  the synthesized attosecond pulse width on the squeezing parameter of the squeezed light. Our findings provide a robust tool for obtaining phase-locked harmonics throughout the HHG spectrum for synthesizing short attosecond pulses.

\end{abstract}
\maketitle

\section{Introduction}

Attosecond pulses are central to the observation and investigation of transient physical phenomena of attosecond time-resolved valence and core-electron dynamics \cite{Krausz2009RMP, Corkum2009NP, Kapteyn2007Science, Chini2014NP}. Presently, attosecond pulses are mainly synthesized from high-order harmonic generation (HHG) of an atom subjected to a strong laser field \cite{Hentschel2001Nature,Paul2001Science}. The corresponding HHG spectrum, having a plateau structure followed by a cutoff, provides a broad spectral bandwidth from extreme ultraviolet to soft x-rays for synthesizing attosecond pulses \cite{Popmintchev2012Science, Gao2022Optica}. However, to produce an attosecond pulse from the HHG, different harmonics should also be phase-locked \cite{Siegmann1986Laser, Antoine1996PRL, Gaarde2002PRL}. Unfortunately, it has been found that the phase locking occurs in the cutoff region, while the harmonic phase
in the broad plateau region of the HHG spectrum is random \cite{Antoine1996PRL, Balcou1996PRA}. This discrepancy results from the fact that the harmonics in the plateau region are contributed by two different (``short" and ``long") electron trajectories \cite{Lewenstein1994PRA,Lewenstein1995PRA, Antoine1996PRL}, whereas these trajectories approximately merge into a single trajectory in the cutoff region. Therefore, only the phase-locked harmonics in the cutoff region are suitable for attosecond pulse synthesis, limiting the ability to achieve short attosecond pulses.

To synthesize a shorter attosecond pulse from the HHG, it is necessary to lock the phase of the harmonics over a wider spectral range. One valid approach is to select either the ``short'' or ``long'' trajectory harmonics in the plateau region. Currently, this single-trajectory harmonic selection can be achieved using a macroscopic phase-matching technique based on the coherent superposition of different atomic harmonics \cite{Lewenstein1995PRA, Salieres1995PRL}, and it has also been demonstrated that a relatively short attosecond pulse can be produced with this technique \cite{Antoine1996PRL,Gaarde2002PRL,Mairesse2003Science,Sansone2006Science, Mairesse2004PRL}. However, it is important to note that the phase-matching conditions are closely related to both the laser intensity and the harmonic order \cite{Lewenstein1995PRA, Salieres1995PRL}. For any given laser parameter, only a fraction of the harmonics within the plateau satisfy the phase-matching conditions for generating single-trajectory harmonics, which restricts obtaining more phase-locked harmonics. On the other hand, another scheme has been proposed to control the ``short'' and ``long'' trajectories at the single-atom level using an orthogonally polarized two-color field \cite{Kim2005PRL,Brugnera2011PRL,Kim2005PRA,Kim2006JPB}. In this method, the electron trajectories are steered in the polarization plane to either return to the core or not, thereby allowing control over the relative contribution of the short and long trajectories. However, this method of manipulating electron trajectories is also most effective for some harmonics. Thus, achieving a broader or even the entire phase-locked HHG spectrum for generating a shorter attosecond pulse remains an ongoing challenge.

In this work, we demonstrate a novel scheme that can realize a short attosecond pulse synthesis from the entire phase-locked HHG spectrum of an atom driven by intense squeezed light. In quantum optics, squeezed light is a typical non-classical state of light \cite{QuantumOptics}. With the recent development of laser technology, strong squeezed lights have become experimentally accessible \cite{Qu1992OC,IskhakovOL2012,PerezOL2014,Finger2015PRL}, and their intensities are gradually approaching the requirements of strong-field physics. This progress of the laser technology has sparked interest in exploring the interaction between quantum lights and atoms \cite{Khalaf2023SA,Gorlach2023NP,Tzur2023NP,Fang2023PRL,Wang2023PRA,Tzur2024LSA}. Here, we use intense squeezed light to interact with an atom to study the attosecond pulse synthesis from the HHG. Our result shows that the harmonics in the whole HHG spectrum can be phase-locked and, by using these harmonics, the width of the attosecond pulse synthesized from these phase-locked harmonics is greatly reduced. The detailed analysis shows that the phase-locking of the harmonics is attributed to the suppression of the long-trajectory contribution and the underlying physics is revealed by developing strong-field approximation (SFA) theory in squeezed light. Furthermore, we uncover the dependence of the attosecond pulse width on the squeezing parameter of the squeezed light. Finally, the experimental feasibility of our scheme for generating short attosecond pulses in intense squeezed light is discussed.

This paper is organized as follows. In Sec. II, we briefly introduce the theoretical method for the HHG of an atom in intense squeezed light. In Sec. III, we present the HHG spectra and the corresponding  attosecond pulses. Subsequently, we reveal the influence of the squeezed light on the length of the attoscond pulse. Finally, our conclusion is given in Sec. IV. Throughout this paper, atomic units (a.u.) are used unless explicitly stated otherwise.

\section{Theoretical methods}

In theory, the interaction of an atom with quantum light is described by the following fully-quantum time-dependent Schr\"odinger equation (TDSE) \cite{QuantumOptics}
\begin{equation}\label{eq1}
i\frac{\partial\rho(t)}{\partial t}=[\left(\hat{H}_0-\textbf{r}\cdot\hat{\textbf{E}} +\hat{H}_f\right), \rho(t)],
\end{equation}
where  $\rho(t)$ is the density matrix of the quantum light-atom system, $\hat{H}_0=\textbf{p}^2/2+U(\textbf{r})$ is the Hamiltonian of the atomic system with the Coulomb potential $U(\textbf{r})$, $\hat{H}_f=
\omega\hat{a}^\dagger\hat{a}$ is the quantized electromagnetic field Hamiltonian with the creation operator $\hat{a}^\dagger$ and the annihilation operator $\hat{a}$, $\hat{\textbf{E}}=i\sqrt{\dfrac{\omega}{2\epsilon_0V}}\left(\hat{a}- \hat{a}^\dagger \right)$ is the quantized electric field operator with vacuum permittivity $\epsilon_0$ and quantum normalization volume $V$, and $-\textbf{r}\cdot\hat{\textbf{E}}$ is the interaction term in the dipole approximation. The initial density matrix $\rho_0=\left|\phi\right>\left<\phi\right| \otimes\rho_{\text{light}}$, where $\left|\phi\right>$ denotes the ground state of the atom and $\rho_{\text{light}}$ is the density matrix of the quantum light field. Using the positive $P$ representation, $\rho_{\text{light}}$ can be expanded as follows \cite{Drummond1980JPA,Kim1989PRA}:
\begin{equation}\label{eq2}
\rho_{\text{light}}=\int P(\alpha,\beta^*)\frac{\left| \alpha\right\rangle \left\langle \beta\right| }{ \left\langle \beta |\alpha\right\rangle}d^2\alpha d^2\beta,
\end{equation}
where $\left| \alpha\right\rangle $ and $\left| \beta\right\rangle $ represent the coherent states of light and $P(\alpha,\beta^*)=\frac{1}{4\pi}\text{exp}\left(-\frac{\left|\alpha-\beta\right|^2}{4}\right)Q\left(\frac{\alpha+\beta}{2}\right)$ with a Husimi distribution $Q(\alpha)$ of quantum light state.

Next, we solve Eq.~(\ref{eq1}) for the evolution of the electron density matrix using the linearity of the density matrix equation. After inserting Eq.~(\ref{eq2}) into Eq.~(\ref{eq1}) and considering the relation $E_{\alpha}=\sqrt{\frac{2\omega}{\epsilon_0V}}\alpha$,
the density matrix of the electron at time $t$ is given by \cite{Gorlach2023NP,Tzur2023NP}:
\begin{equation}\label{eq5}
\rho_{\text{e}}(t)=\int d^2E_\alpha \tilde{Q}(E_\alpha)\left|\phi_{E_\alpha}(t) \right\rangle\left\langle \phi_{E_\alpha}(t)\right|,
\end{equation}
where $\tilde{Q}(E_\alpha)=\lim\limits_{V\to\infty}\frac{\epsilon_0V}{2\omega}Q\left(\sqrt{\frac{\epsilon_0V}{2\omega}}E_\alpha\right)$ denotes an electric-field quasiprobability distribution of quantum light and $\left|\phi_{E_{\alpha}}(t) \right\rangle$ represents the electron wave function in coherent-state (classical) light. Here, $\left|\phi_{E_{\alpha}}(t) \right\rangle$ is obtained by solving the TDSE:
\begin{equation}\label{TDSE}
i\frac{\partial}{\partial t}\left|\phi_{E_{\alpha}}\left(t\right)\right>=[\hat{H}_0-\textbf{r}\cdot \textbf{E}_{\alpha}(t)]\left|\phi_{E_{\alpha}}\left(t\right)\right>,
\end{equation}
where $\textbf{E}_{\alpha}(t)$ is the classical field component of the coherent state $\left| \alpha\right\rangle =\left| \alpha_x+i\alpha_y\right\rangle$ and is expressed as
$\textbf{E}_{\alpha}(t)=\left\langle\alpha\right|\hat{\textbf{E}}(t) \left| \alpha \right\rangle\hat{z}=\left[-E_{\alpha_x}\sin(\omega t)+E_{\alpha_y}\cos(\omega t)\right]\hat{z}$ along the $\hat{z}$ direction.
In this work, we use a phase-squeezed coherent state light $\left|\gamma, r \right\rangle$ with squeezing parameter $r$. According to the nature of the Husimi distribution of the phase-squeezed coherent state \cite{Kim1989PRA}, the electric field quasiprobability distribution $\tilde{Q}(E_\alpha)$ is given by
\begin{equation}\label{eq7}
\tilde{Q}(E_\alpha)=\frac{1}{\sqrt{2\pi\left|E_\text{vf} \right|^2} }  e^{-\frac{(E_{\alpha_y}-E_{\gamma_{y}})^2}{2\left|E_\text{vf}\right| ^2} } \delta(E_{\alpha_x}-E_{\gamma_{x}}),
\end{equation}
where $|E_\text{vf}|^2=\frac{2\omega}{\epsilon_0V}\sinh^2(r)$ is the electric field fluctuation amplitude of the squeezed light.

To derive the HHG from the electron density matrix, we calculate the expectation value of the dipole moment $D(t)$ \cite{QuantumOptics}:
$
D(t)=\text{Tr}[z\rho_e(t)]=\int d^2E_\alpha\tilde{Q}(E_\alpha)D_{E_\alpha}(t) ,
$
where $D_{E_\alpha}(t)=\left\langle \phi_{E_\alpha}(t)|z|\phi_{E_\alpha}(t)\right\rangle$ denotes the expectation value of the dipole moment for a coherent state $\left| \alpha\right\rangle$.
Then, the $n$th-order harmonic can be obtained from the Fourier transform of the dipole moment:
\begin{align}\label{HHG}
D(n\omega)= \frac{\omega^2}{t_f-t_i}\int_{t_i}^{t_f} D(t) e^{-i n \omega t} dt   = \int d^2 E_{\alpha}\tilde{Q}(E_{\alpha}) D_{E_{\alpha}}(n\omega),
\end{align}
where $D_{E_{\alpha}}(n\omega)=\frac{\omega^2}{t_f-t_i}\int_{t_i}^{t_f} D_{E_{\alpha}}(t) e^{-i n \omega t} dt$ is the amplitude of the $n$th-order harmonic in a coherent light $\left| \alpha\right\rangle$.
Eq.~(\ref{HHG}) indicates that the final harmonic amplitude in the squeezed light is equal to the superposition of the harmonics from different coherent lights. After obtaining the HHG amplitude, we can calculate the attosecond pulse by superimposing harmonics in any given energy range:
\begin{equation}\label{eq9}
I(t)=\left|\sum_n D(n\omega) e^{in\omega t} \right| ^2.
\end{equation}
In practice, a one-dimensional model atom with a soft-core Coulomb potential $U(x)=-1/\sqrt{x^2+a}$ is used for simplicity. When the soft-core parameter $a=0.482$, its ionization potential $I_p$ is about 24.59 eV, which corresponds to the ground state of the He atom.	

\section{Results and discussion}
\begin{figure}[t]
	\centering
	\includegraphics[scale=0.4]{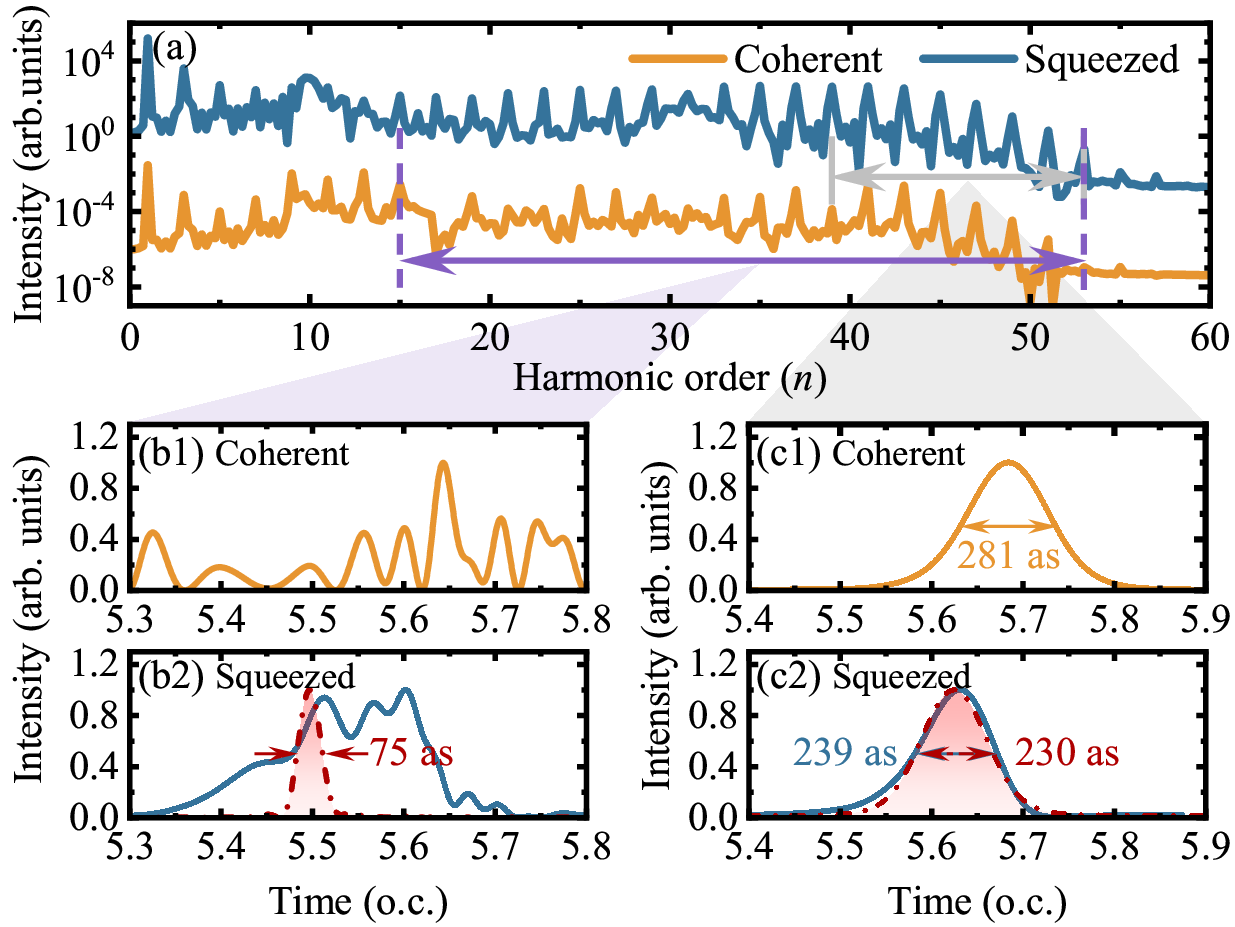}
	\caption{(a) HHG spectra obtained with a coherent light $|\gamma\rangle$ (solid orange curve) and a squeezed light $\left|\gamma, r \right\rangle $ (solid blue curve). (b1) and (c1) Normalized temporal profile of attosecond pulses synthesized from the 15th to 53rd and the 39th to 53rd order harmonics for the coherent light. In (b2) and (c2), the solid blue curves are the same as that in (b1) and (c1), but for the squeezed light, while the red dashed-dotted curves denote the synthesized attosecond pulses after removing the quadratic component of the harmonic phases. For more details, see the text. In our simulation, the $\gamma$ is chosen to correspond to an average light intensity of about $I=2\times10^{14}$ W/cm$^2$, corresponding to the peak electric field $E_0=|E_{\gamma_x+i\gamma_y}|=0.0756$ a.u. with $E_{\gamma_{x}}=0$ a.u. and $E_{\gamma_{y}}=0.0756$ a.u. The squeezing parameter $r$ is given by $\sinh^2(r)=0.0016|\gamma|^2$ and accordingly, the electric field fluctuation amplitude $E_\text{vf}=0.04 E_{0}$. The laser field with a wavelength $\lambda=800$ nm has a trapezoidal profile, ramping up and down over two cycles and remaining constant over six cycles.}
	\label{imag1}
\end{figure}

Figure~\ref{imag1}(a) presents the HHG spectra obtained with coherent light $|\gamma\rangle$ (solid orange curve) and squeezed light $\left|\gamma, r \right\rangle$ (solid blue curve), respectively. Both spectra show a plateau structure with a sharp cutoff around $I_p+3.17U_p\approx40\omega$, where $U_p$ is the ponderomotive energy \cite{Krause1992PRL,Schafer1993PRL,Corkum1993PRL}. Using the obtained harmonics, we first illustrate the attosecond pulses synthesized from the 39th to 53rd order harmonics in the cutoff region (marked with gray arrow) in Figs.~\ref{imag1}(c1) and (c2). For the coherent light, the  pulse width is about 281 attoseconds (as), while it is slightly reduced to about 239 as for the squeezed light [see the blue solid curve in (c2)]. On the other hand, we use the whole HHG above threshold ($n> I_p/\omega \approx 15$  \cite{Krause1992PRL,Schafer1993PRL,Corkum1993PRL} marked with purple arrow), and  the corresponding attosecond pulses are shown in Figs.~\ref{imag1}(b1) and (b2). For coherent light, the temporal profile is rather broad with several peaks and thus no well-defined attosecond pulse can be obtained, which is consistent with previous results \cite{Antoine1996PRL,Gaarde2002PRL}. In contrast, for the squeezed light, an attosecond pulse with a relatively broad width is generated (see the blue solid curve). In the following, we will understand the effect of the squeezed light on the attosecond pulse synthesis and show how to obtain a shorter attosecond pulse under the squeezed light.

\begin{figure}[t]
	\centering
	\includegraphics[scale=0.26]{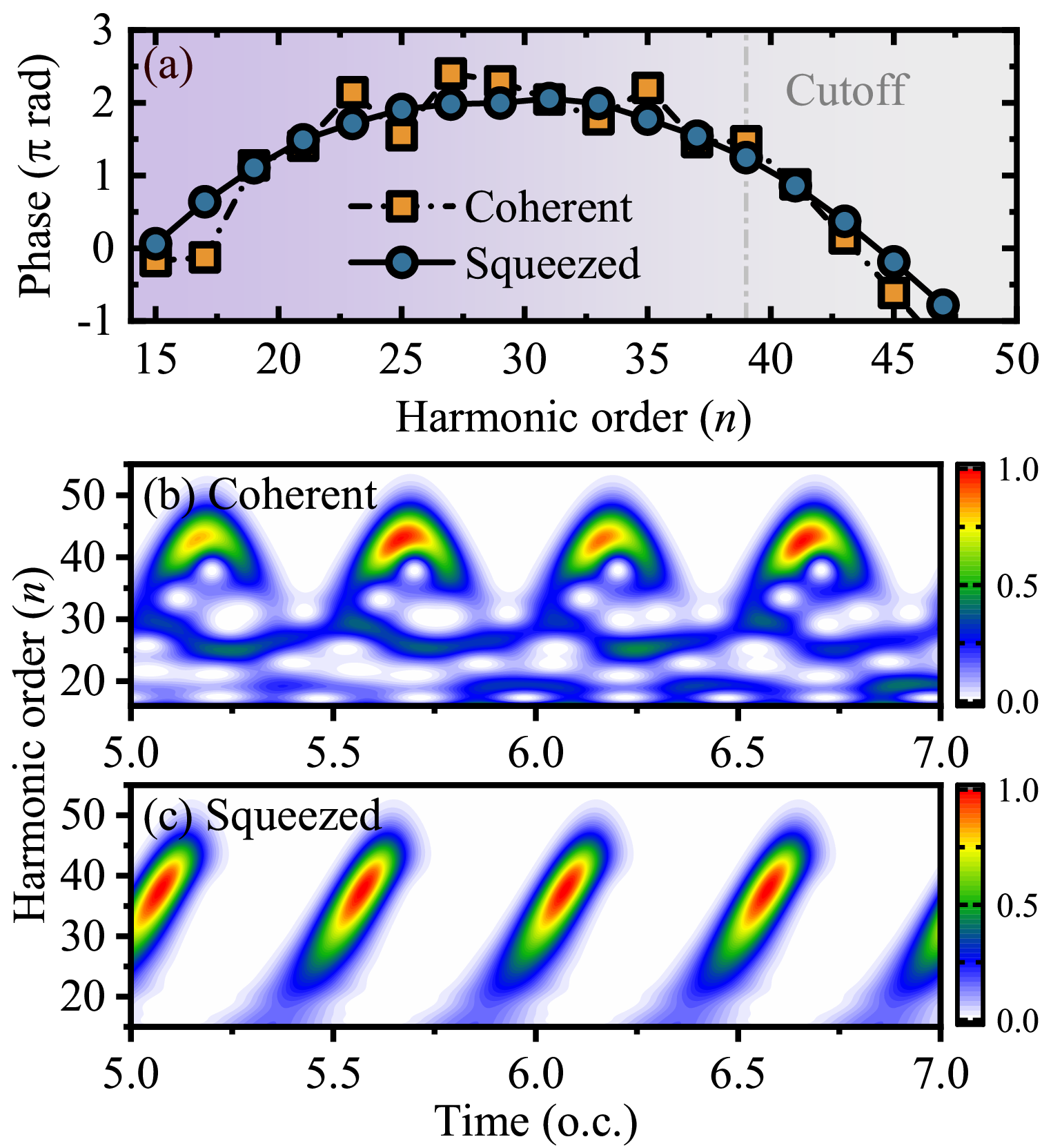}
	\caption{ (a) The harmonic phases as a function of the harmonic orders in the coherent light (orange orthogon) and squeezed light (blue circle), respectively. (b) and (c) The corresponding time-frequency spectra of the HHG.}
	\label{imag2}
\end{figure}

Considering that the temporal structure of attosecond pulses is affected by the harmonic phase \cite{Gaarde2002PRL}, we show in Fig.~\ref{imag2}(a) the phases of the HHG obtained with coherent and squeezed lights, respectively. For coherent light, the phase changes smoothly with harmonic order in the cutoff region, while the phase change in the plateau region is random. A similar result can be found in Ref.~\cite{Antoine1996PRL}. Thus, only the harmonics in the cutoff region can be used to produce the attosecond pulse [see Fig.~\ref{imag1}(c1)]. In contrast, for squeezed light, the harmonic phases change smoothly in the whole HHG spectra and show approximately a simple quadratic structure. Thus, all harmonics are phase-locked \cite{Schafer1997PRL} and can be used to synthesize attosecond pulses [see the blue solid curve in Fig.~\ref{imag1}(b2)]. Surprisingly, however, even after considering the broader spectral bandwidth, the pulse width in Fig.~\ref{imag1}(b2) is still larger than that in Fig.~\ref{imag1}(c2). This abnormal result is due to a linear frequency chirp in time caused by the quadratic spectral phase variation \cite{Mairesse2003Science,Kim2007PRL,Schafer1997PRL}.
Fortunately, this quadratic component can now be completely removed experimentally by manipulating the chirp \cite{Schafer1997PRL,Kim2007PRL,Sansone2006Science,Bouchet2012NJP}. Thus, we recalculate the attosecond pulses in squeezed light after removing the quadratic component of the harmonic phases (see the red dashed-dotted curves in Figs.~\ref{imag1}(b2) and (c2)). In this case, the pulse width decreases slightly to 230 as in (c2), while it is greatly reduced to 75 as in (b2).

To understand the phase change for different laser fields in Fig.~\ref{imag2}(a), we perform the time-frequency analysis of the resulting HHG \cite{Carrera2006PRA}. For the coherent light  in Fig.~\ref{imag2}(b), there are arc structures  from  the short and long trajectories \cite{Carrera2006PRA}.
In addition, there is a relatively weak structure, which arises from multi-return trajectories \cite{Li2015NC}. The interference  of these trajectories leads to the random change of the harmonic phase in the plateau region shown in Fig.~\ref{imag2}(a) \cite{Antoine1996PRL,Balcou1996PRA,Lewenstein1995PRA}.
However, for the squeezed light in Fig.~\ref{imag2}(c), only the left part of the arc can be observed, while the right part of the arc structure and other weak structures are suppressed. Thus, the short-trajectory contribution becomes dominant, and accordingly, the phase distribution shows a quadratic structure \cite{Mairesse2003Science,Kim2007PRL,Schafer1997PRL}, just as shown in Fig.~\ref{imag2}(a).

Furthermore, to gain insight into the suppression of the long-trajectory harmonics in squeezed light, we turn to the SFA theory \cite{2002PRA}. According to the SFA theory, the harmonic amplitude in coherent light can be written mainly as the superposition of the contributions from the short and long electron trajectories: $D_{E_\alpha}(n \omega)=\sum_k a^k_{E_\alpha} e^{i S^k_{E_\alpha}(\textbf{p},t,t')}$, where $k$ denotes different trajectory with weight $a^k_{E_\alpha}$ and phase $S^k_{E_\alpha}(\textbf{p},t, t')$. This phase is equal to the action accumulated by the electron along the corresponding trajectory. For a given harmonic order, the values of the canonical momentum $\textbf{p}$, the ionization time $t$, and the reconbination time $t'$ of each trajectory can be obtained by solving saddle-point equations \cite{2002PRA,2005PRA}. Substituting $D_{E_\alpha}(n \omega)$ into Eq.~(\ref{HHG}), we obtain the final harmonic amplitude in squeezed light:
\begin{equation}\label{SFA}
D(n \omega)=\sum_k\int d^2E_{\alpha}\tilde{Q}(E_\alpha) a^k_{E_{\alpha}} e^{i S^k_{E_{\alpha}}(\textbf{p},t,t')}.
\end{equation}
According to this equation, the final amplitude of the long or short trajectories is affected by the phase distribution with respect to the electric field $E_{\alpha}$.

\begin{figure}[t]
	\centering
	\includegraphics[scale=0.33]{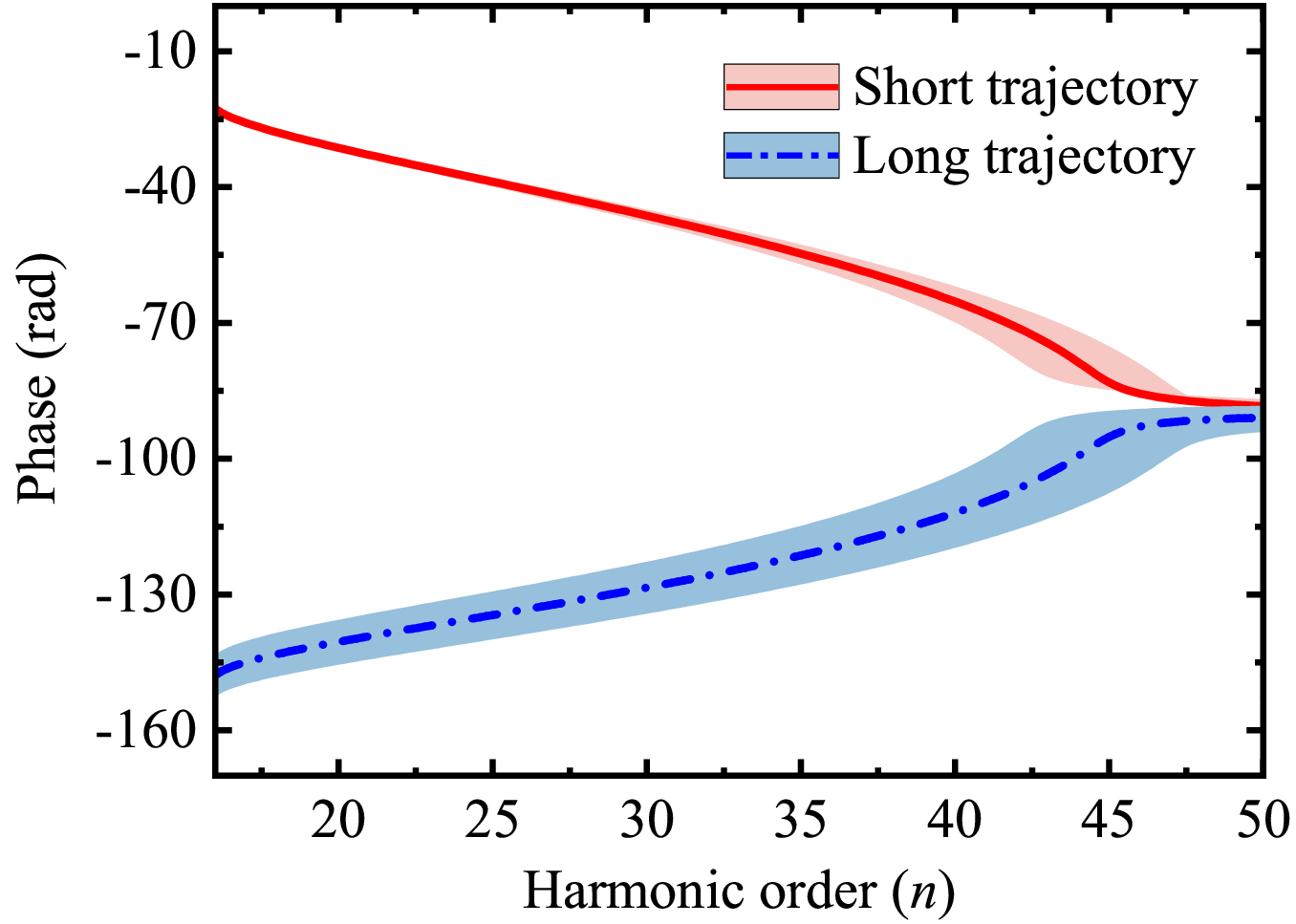}
	\caption{ Phase of the long and short electron trajectories for HHG. The solid red curve and the dashed-dotted blue curve denote the harmonic phase in a coherent light with a determined electric field $E_0$. The shaded regions correspond to the phase distribution caused by the electric field quasiprobability distribution $\tilde{Q}(E_\alpha)$ of the squeezed light.}
	\label{imag3}
\end{figure}

In Fig.~\ref{imag3} we present the phase of the short and long trajectories as a function of the harmonic order in the coherent and squeezed lights, respectively. For coherent light with a determined electric field $E_0$, the phase of the short and long trajectories is described by the solid red curve and the dashed-dotted blue curve. However, for squeezed light, due to the electric field quasiprobability distribution of $\tilde{Q}(E_{\alpha})$ in Eq.~(\ref{SFA}), there is a phase distribution for both short and long trajectories. Here, we consider the full width at half maximum (FWHM) of the electric field distribution and the corresponding phase distribution is marked by the shaded regions in Fig.~\ref{imag3}. It is interesting to note that the phase distribution of the long trajectory becomes broad, while for the short trajectories, it is rather small, especially in the plateau region. The reason is that the electron trajectory phase can be approximated by \cite{Salieres2001Science,Auguste2009PRA}: $S_{E_\alpha}(\textbf{p},t, t')\approx  -\tau E_{\alpha}^2 /4\omega^2$,  where $\tau$ is the excursion time. For the long trajectory with long $\tau$, its phase is sensitive to the change of the electric field, leading to the broad phase distribution. In contrast, for the short trajectory, the value of $\tau$ is small, resulting in the small phase distribution. As a result, the harmonic amplitude of the long trajectories with broad phase distribution is suppressed due to destructive interference. This result is only related to the nature of the electron trajectory and is therefore independent on the target system. Similarly, for the multi-return trajectories with longer excursion time \cite{Li2015NC}, their harmonic amplitudes are also suppressed in squeezed light.

\begin{figure}[t]
	\centering
	\includegraphics[scale=0.32]{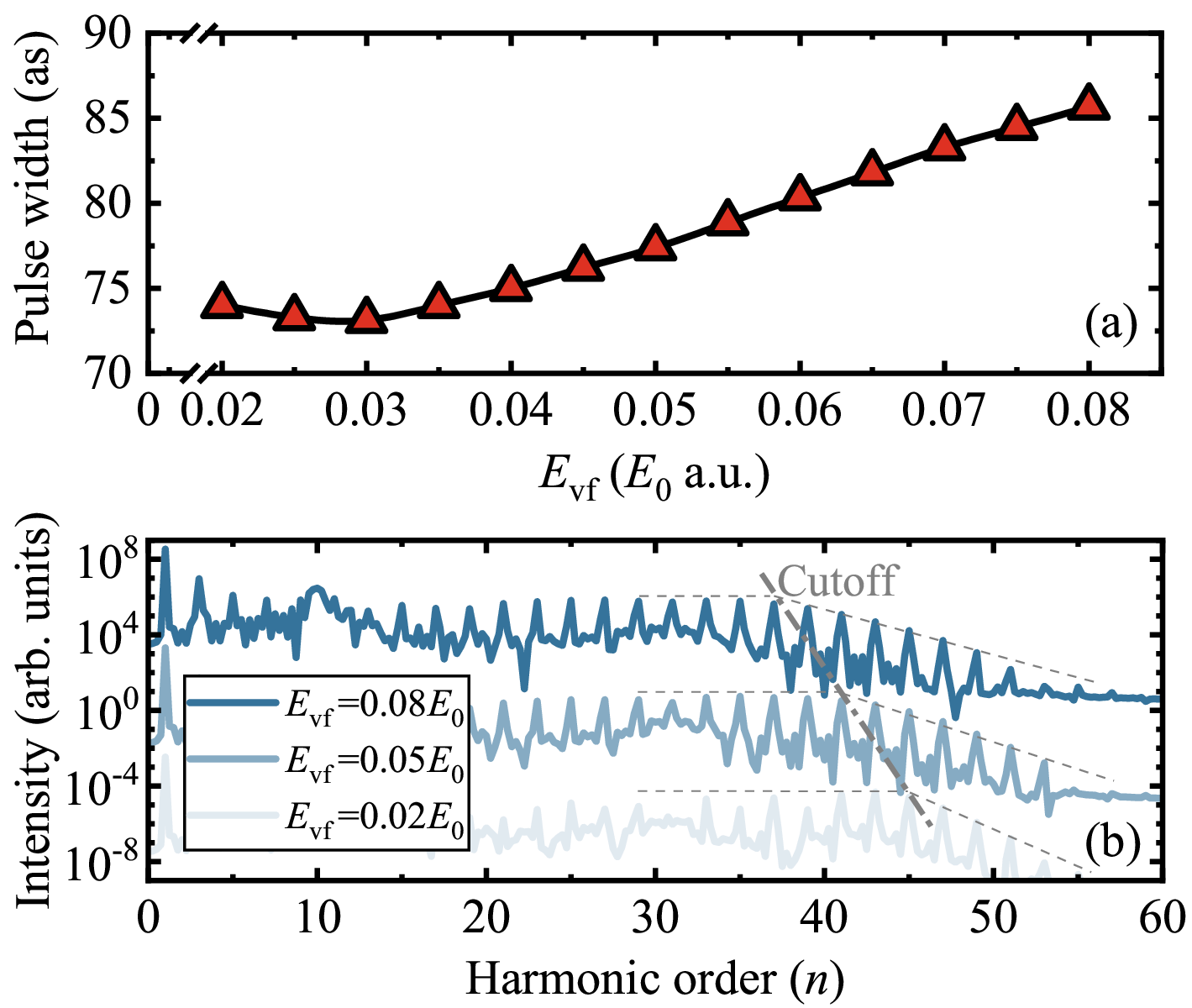}
	\caption{ (a) The dependence of the attosecond pulse width on the electric field fluctuation amplitude $E_\text{vf}$ of the squeezed light. The attosecond pulse is synthesised from 15th to 53rd order harmonics after removing the quadratic component of the harmonic phases.  (b) The HHG spectra obtained with different $E_\text{vf}$.}
	\label{imag4}
\end{figure}

Next, we investigate the dependence of the attosecond pulse width on the electric field fluctuation amplitude $E_\text{vf}$ of the squeezed light in Eq.~(\ref{eq7}). Figure~\ref{imag4}(a) shows the width of the attosecond pulse as a function of $E_\text{vf}$. As $E_\text{vf}$ increases, the width of the attosecond pulse first decreases slightly and then increases.  The smallest width can reach to 73 as, when $E_\text{vf}$ is about $0.03E_0$. The reason for the non-monotonic changes of the width is that initially the increase of $E_\text{vf}$ can more effectively suppress the long-trajectory contribution, leading to the generation of a shorter attosecond pulse. However, as $E_\text{vf}$ is further increased, the phase distribution of the short trajectories near the cutoff also becomes broader; see Fig.~\ref{imag3}. Due to destructive interference, the corresponding harmonic amplitude is suppressed [see Fig.~\ref{imag4}(b)]. Therefore, the spectral bandwidth available for synthesizing attosecond pulses becomes narrow, resulting in an increase in the pulse width.

Finally, we discuss the experimental feasibility of our scheme for synthesizing short attosecond pulses from HHG in squeezed light. In our scheme, intense squeezed light is required for HHG. Currently, a relatively weak squeezed light can be generated by combining a strong coherent beam with a bright squeezed vacuum beam \cite{QuantumOptics,Paris1996PL}. To obtain  intense squeezed light used in our work, the intensity of the bright squeezed vacuum beam should be increased. For intense squeezed light with, e.g., $E_\text{vf}=0.03E_0$, the intensity of the bright squeezed vacuum beam is about $I_\text{vf}= \frac{1}{2}\epsilon_0cE_\text{vf}^2=9\times10^{-4} I_0=1.8\times10^{11}$ W/cm$^2$ \cite{Tzur2023NP}. Currently, a bright squeezed vacuum femtosecond beam with an energy $\mathcal{E}$ of 350 nJ, a width of $\tau_p=$140 fs and a spot diameter of $d=$18.5 $\mu$m can be achieved experimentally \cite{Finger2015PRL}. Accordingly, its intensity is about $4\mathcal{E}/(\pi\tau_p d^2)= 9.3\times10^{11}$ W/cm$^2$ \cite{Chang2016}, which is larger than $I_\text{vf}$. Therefore, the intense squeezed light that meets the requirements of our scheme could be generated under the current experimental conditions.

\section{Conclusion}

In summary, we have studied the attosecond pulse synthesis from the HHG in intense squeezed light by solving fully-quantum TDSE. Our result shows that the harmonics in the whole HHG spectrum can be phase-locked, and the width of the attosecond pulse synthesized from these phase-locked harmonics is greatly reduced. The time-frequency analysis of the HHG reveals that  the phase-locking of the harmonics is due to the suppression of the long-trajectory harmonics in the whole spectrum. By developing the SFA theory in squeezed light, the suppression of the long-trajectory contribution is attributed to the destructive interference of the trajectories with a broad phase distribution in intense squeezed light, and this process is independent on the target system. Thus, our  scheme, as demonstrated in this work, provides a robust tool for obtaining phase-locked harmonics throughout the HHG spectrum for synthesizing short attosecond pulses, and could be extended to other attosecond pulse generation systems, such as attosecond pulse generation from solid-state HHG \cite{Li2020NC}.

\section*{Acknowledgments}
We thank Dr. Cheng Gong and Dr. Lu Wang for helpful discussion.
This work is supported by the National Key Program for S$\&$T Research and Development (No. 2019YFA0307702), the National Natural Science Foundation of China (Nos. 11922413, 12121004, and 12274420),  and CAS Project for Young Scientists in Basic Research (No. YSBR-055).

\end{document}